\documentstyle[aaspp]{article}
\input{psfig}
\tighten

\begin{document}
\def\etal{{\it et al.\/}}
\def\cf{{\it cf.\/}}
\def\ie{{\it i.e.\/}}
\def\eg{{\it e.g.\/}}

\title{Lense-Thirring Precession and QPOs in Low Mass X-Ray Binaries} 
\author{{\bf Luigi Stella\altaffilmark{1,3} and Mario Vietri\altaffilmark{2}}} 
\altaffiltext{1}{Osservatorio Astronomico di Roma, Via dell'Osservatorio 2, 
00040 Monte Porzio Catone (Roma), Italy, 
e-mail: stella@coma.mporzio.astro.it} 
\altaffiltext{2}{Universit\`a di Roma 3, Via della Vasca Navale 84, 
00147 Roma, Italy, e-mail: vietri@corelli.fis.uniroma3.it }
\altaffiltext{3}{Affiliated to the International Center for Relativistic
Astrophysics}

\begin{abstract}
We show in this Letter that relativistic dragging of 
inertial frames around fast rotating collapsed stars is 
substantial and can give rise to  observable effects. We apply this to 
the kHz quasi periodic oscillations (QPOs) sources, 
low mass X-ray binaries (LMXRBs) containing an accreting 
neutron star. Within the beat frequency model, both the Keplerian frequency
of the innermost region of the accretion disk ($\sim 0.3-1.2$~kHz) and the 
neutron star spin frequency ($\sim 0.3-0.4$~kHz) are directly observed. 
From these the Lense-Thirring precession frequency (tens of Hz)
of the same material in the innermost disk regions which gives rise to the 
kHz QPOs is determined within a factor of $\sim 4$, depending on the neutron 
star equation of state. The classical contribution from neutron star oblateness 
decreases the precession frequency slightly. The broad peaks at frequencies 
$\sim 20-40$~Hz in the power spectra of the ``Atoll"-sources 
4U~1728-34, 4U~0614+091 and KS~1731-260 and their variations 
with the higher kHz QPO frequency are well matched by 
Lense-Thirring precession of material in the innermost disk region. 
We also suggest that the $\sim 15-50$~Hz horizontal branch QPOs 
of GX~5-1 and GX~17+2 (and likewise other ``Z"-type 
low mass X-ray binaries) arise from the same mechanism. 

\end{abstract}

\section{Introduction}

The variety of QPO modes from old accreting neutron
stars in low mass X-ray binaries (LMXRBs) 
(see \eg\/ van der Klis 1995) has been recently enriched by 
the  discovery of kiloHertz oscillations in light curves obtained with the 
Rossi X-ray Timing Explorer (RXTE). These are the first QPOs that involve 
timescales comparable to the dynamical timescales in the vicinity of a 
collapsed star. There are currently eleven LMXRBs 
from which QPOs have been detected between frequencies of $\sim 0.3$ and 
$1.2$~kHz. 

Most of these are type I X-ray bursters with
persistent  luminosities in the $10^{36}-10^{37}$~ergs~s$^{-1}$ range, that
according  to the classification of Hasinger and van der Klis (1989) either 
belong to (4U~1728-34, 4U~1608-52, 4U~1636-536, 4U~1735-444 and 4U~1820-30) 
or are suspected members of (4U~0614+091 and KS~1731-260) the  ``Atoll"
group. Sco~X-1, GX~5-1 and GX~17+2 are instead high luminosity ``Z"-type LMXRBs.

A pair of kHz QPO peaks is often  present
in the power spectra.  This is the case for the persistent emission of
the five Atoll sources,  
at least on occasions and/or over some range of count rates. In 
4U~1728-34 and 4U~0614+091 the centroid frequency of the QPOs is positively
correlated with the source count rate 
across the factor of $\sim 2$ QPO frequency variations (from $\sim
500$ to $\sim 1100$~Hz) observed from both sources.  When a pair
of peaks is present, the difference $\Delta\nu$ between their
centroid  frequencies $\nu_1$ and $\nu_2$ is consistent with being constant. 
In 4U~1728-34 $\Delta\nu$ remains around 
$\simeq 363$~Hz (within a few percent)
as the frequency of the kHz QPOs changes by $\sim 20$\%. 
A nearly coherent signal at $\sim 363$~Hz
has been detected during a number of X-ray bursts (Strohmayer \etal\/ 1996a). 
In the persistent emission of 4U~0614+091 the separation $\Delta\nu$
is consistent with $\sim  323$~Hz (within $\sim 10$\%),
across a $\sim 50$\% variation of the QPO frequency and a $\sim 100$~d 
interval between two  RXTE observations. 
During part of an observation, a third narrower
peak has been marginally detected at a frequency of $\sim 328$~Hz, consistent
with the separation of the pair of kHz QPO peaks 
(Ford \etal\/ 1997). In the persistent emission of KS~1731-260, two kHz 
QPO peaks have been detected around $\sim 900$ and $\sim 1160$~Hz. The 
frequency difference ($\sim 260$~Hz) between these QPOs is consistent with 
half the frequency of the nearly coherent $524$~Hz signal observed in a 
type I burst from this source (Smith \etal\/ 1997; Wijnands \& van der Klis 
1997). kHz QPOs have also been discovered in 
4U~1608-52 (Berger \etal\/ 1996) and 4U~1735-444 (Wijnands \etal\/ 1996). 

A QPO peak with an intensity-dependent frequency 
from $\sim 550$ to $\sim 800$~Hz has been detected in 
4U~1820-30. Close to the upper end of this range a second QPO peak at 
$\sim 1070$~Hz is present (Smale, Zhang \& White 1997). In the 
persistent emission of 4U~1636-536 a pair of QPO peaks around 840-920~Hz and
1150-1200~Hz have been detected, the frequency separation of which is 
constant around $\Delta\nu\sim 255\pm 25$~Hz. 
A nearly coherent signal at $\sim 581$~Hz
modulates  the X-ray flux of this source during type I bursts (Zhang \etal
1996, Wijnands \etal 1997); note that half this frequency is 
compatible with the separation $\Delta\nu$.  
Periodic oscillations at 589~Hz have also been observed during type I bursts 
from an unknown burster in the direction of the galactic centre, perhaps 
MXB~1743-29 (Strohmayer \etal\/ 1996b). 

GX~5-1, GX~17+2 and Sco~X-1, currently the only high-luminosity 
``Z"-type LMXRBs
in the group, display kHz QPOs ($\sim 0.3-1.1$~kHz) with a centroid frequency 
which is positively correlated with the inferred mass accretion rate. 
In GX~5-1 a pair of kHz QPO peaks has been detected over part of the
horizontal branch (HB), which mantain a constant separation of 
$\Delta\nu \simeq 325$~Hz. The higher frequency QPO peak
of GX~5-1 increases from $\nu_2\sim 570$ to $\sim 900$~Hz as the source moves
from the  left to the right end of the HB 
(van der Klis \etal\/ 1996a). 
In GX~17+2 a kHz QPO peak is observed to drift from 
$\sim 680$ to $880$~Hz QPO as the source moves down the upper part of the
so-called normal branch. A higher frequency peak separated by $\sim 306$~Hz
is observed close to the lower end of this range, together with an additional
QPO peak around $\sim 60$~Hz, interpreted as HB oscillations  
(van der Klis \etal\/ 1997a).  
In Sco~X-1, $\sim 0.5-1.1$~kHz QPOs have been detected 
in the normal and flaring branches; their frequency correlates 
well both with the inferred variations of the mass accretion rate 
and the frequency variations of the $\sim 6-20$~Hz QPO that
occur along the two branches.  
A pair of kHz QPO peaks has been detected on occasions. However
their separation decreased from $\Delta\nu \sim 310$ to $\sim 230$~Hz
as the higher frequency peak moved from $\sim 875$ to 
$\sim 1085$~Hz. Correspondingly 
additional QPOs at $\sim 40-50$~Hz were present, which are intepreted as 
HB QPOs (van der Klis \etal\/ 1996b, 1997b). 

Modelling of kHz QPOs has focussed on 
the presence of peaks at $\nu_1$ and $\nu_2$, 
the constancy of their separation $\Delta\nu$ 
and the coherence of the signal (either during  type I bursts or in the 
persistent emission) at frequency $\Delta\nu$ (or $2\Delta\nu$)\footnote{ 
Though these properties are not observed in all LMXRBs with kHz QPOs, they
appear to be common, with Sco~X-1 the only clear 
exception to the constancy of $\Delta\nu$ (White \& Zhang 1997)}. 

The application of beat frequency concepts to these signals is
natural. In the magnetospheric beat frequency model, MBFM, 
originally developed to interpret the 5-50~Hz QPOs from high luminosity 
``Z"-type LMXRBs (Alpar \& Shaham 1985; Lamb \etal\/ 1985),
matter inhomogeneities in the accretion disk 
cause accretion to be modulated at the  beat
frequency between the Keplerian frequency, $\nu_K$, of the innermost 
region of the accretion disk ($r_{in}$) at   
the magnetospheric boundary, and the  neutron star spin frequency, $\nu_s$. 
As $r_{in}$ decreases with increasing accretion rates, $\nu_K$ increases 
and so does the QPO beat frequency $\nu_{b}= \nu_K-\nu_s$.
Of the various QPO modes with frequencies $< 100$~Hz in 
``Z"-type LMXRBs, only HB QPOs 
are currently interpreted in terms of the MBFM
(for a review see Lamb 1991). 
The basic physical picture of the MBFM has been confirmed 
mainly through the study of QPOs in a few accreting X-ray pulsars
(EXO~2030+375 and A~0535+26 in particular), where, unlike ``Z"-type LMXRBs, 
the neutron star spin frequency $\nu_s$ is (by definition) clearly 
detected, and independent estimates of the neutron star
magnetic field exist (Angelini \etal\/ 1989; Finger \etal\/ 1996). 

Besides the signals at $\nu_s$ and $\nu_b$,  
an additional QPO signal at $\nu_K$ is expected
in the MBFM from inhomogeneities at the inner disk boundary. 
For self-luminous blobs, the amplitude of this signal 
should be higher for smaller values of $r_{in}$,  
due to the larger local gravitational energy
and Keplerian velocities (causing a larger Doppler 
boosting around the orbit). Similarly, a small value of $r_{in}$ increases 
the range of inclination angles for which a signal at $\nu_K$ can be 
caused by the neutron star occulting the blobs at $r_{in}$ or
conversely, by the blobs at $r_{in}$ obscuring part of the radiation 
produced closer to the neutron star surface. 
  
In the MBFM for kHz QPOs from LMXRBs, the 
lower frequency QPOs are interpreted as the beat frequency 
($\nu_1\simeq\nu_b$) and the higher frequency QPOs as the Keplerian frequency 
($\nu_2\simeq\nu_K$) at the magnetospheric boundary. 
The nearly coherent signal, occasionally detected 
at a frequency compatible with the separation 
$\Delta\nu$ of the QPO pair (or with $2\Delta\nu$), provides the first direct
evidence of the neutron star spin frequency $\nu_s$ in 
LMXRBs containing an old neutron star (Strohmayer \etal\/ 1996a; 
Ford \etal\/ 1997). Inferred magnetospheric radii 
$r_m \simeq r_{in} =  (GM)^{1/3}(2\pi\nu_K)^{-2/3} = 
15 M_{0}^{1/3}\nu_{K3}^{-2/3}$~km,
(here $M = M_{0} {\rm M_{\odot}}$ is the neutron star mass and 
$\nu_{K} = 10^3 \nu_{K3}$~Hz) correspond to a couple
of neutron star radii at the most. 
The neutron star spin frequency of several hundred~Hz and the inferred 
magnetic field strength of $\sim 10^8-10^9$~G agree with spin--up by 
accretion torques, and scenarios for the formation of millisecond pulsars. 

Miller \etal\/ (1996) proposed an alternative beat frequency model
based on the idea that the magnetic field of the neutron star is so low 
that it does not affect the dynamics of the innermost disk regions and, yet,
high enough to give rise to beamed radiation at the neutron star surface. 
We do not discuss the details of this model, but notice that
here too the QPO signal at the highest frequency $\nu_2$ corresponds to
the innermost Keplerian frequency of the disk, which in
this model ends at the sonic radius where the radial velocity becomes 
supersonic (as opposed to the magnetospheric radius), and the
lower frequency $\nu_1$ is the beat frequency between the Keplerian
frequency and the neutron star spin frequency, exactly like in the MBFM.

\section{Lense-Thirring Precession in LMXRBs with kHz QPO}

The parameters inferred from beat frequency models for the kHz QPO from LMXRBs
are  such that frame dragging, one of the general relativistic effects of the
``gravitomagnetic" field associated with rotating bodies, is substantial
and can lead to observable phenomena. 
In particular the orbital plane of a test particle,
if not coaligned with the equatorial plane, must
precess in a prograde fashion 
around the angular momentum axis of the rotating object. 
In the weak field limit the nodal precession frequency, first calculated 
by Lense \& Thirring (1918) (see also Wilkins 1972), is given by
$\nu_{LT} = GMa/(\pi c^2 r^3)$,
where $M$ and $a$ are the mass and specific angular momentum of the rotating 
object and $r$ is the radius of the test particle (circular) orbit. 
By using $Ma = 2\pi I \nu_s$, 
and $r = (GM)^{1/3}/ (2\pi \nu_K)^{2/3}$ this can be written as 
\begin{equation}
\label{1}
\nu_{LT}= {{8 \pi^2 I \nu_K^2 \nu_s}\over{c^2 M}} = 13.2\ I_{45}M_0^{-1} 
\nu_{K3}^2 \nu_{s2.5} \ \ \ {\rm Hz}
\end{equation}
where $I = 10^{45} I_{45}$~g~cm$^2$ is the moment of inertia of
the neutron star and $\nu_s = 300 \nu_{s2.5}$~Hz its spin frequency. 
By using the values of $\nu_s$ and $\nu_K$ inferred from the application
of beat frequency models to the kHz QPOs from LMXRBs, it is apparent
that Lense--Thirring frequencies of a few tens of Hz are expected in the
innermost disk region which produces the QPO signal at $\nu_K$.
Note also that in this regime, higher order terms in $a/M$ would provide
a correction to Eq.~(1) at a level of less than one percent.

Since sources with a pair of kHz QPO peaks allow
identification of both $\nu_K$ and $\nu_s$, the only parameter 
in Eq.~(1) that is not immediately
identified from observations is $I/M$. The theoretical
uncertainty in its determination depends on the choice
of the equation of state (EOS) and the star mass:
from models of rotating neutron stars (Friedman, 
Ipser and Parker 1986; Cook, Shapiro \& Teukolsky 1994), it can be seen that 
$I/M$  varies in the range
$0.5 < I_{45}/M_0 < 2$ for any mass and EOS.
We notice that $\nu_{LT}$ in Eq.~(1) is thus independent of any other
uncertainty  related to LMXRBs, such as 
source distance, mass accretion rate or efficiency of conversion of
gravitational energy into radiative energy. 

The neutron star spin frequencies discussed here lead to
a stellar oblateness which, in turn, causes the appearance of a
quadrupole term in the gravitational potential.
It is well--known that the quadrupole term causes a precession of 
orbits tilted away from the star equatorial plane,
with the precession frequency 
$\propto r^{-7/2}$ (as opposed to $r^{-3}$ for the Lense-Thirring
precession).  For the small radii of interest
to us, it is important to check that this classical precession rate 
is still smaller than the Lense--Thirring rate, to make 
sure that the effect we predict is a genuine relativistic one. 
The classical precession frequency $\nu_{cl}$ is given by 
(Kahn and Woltjer 1959)
\begin{equation}
\label{kahn}
\nu_{cl} = \frac{3}{8\pi^2}\frac{\phi_2 \cos\beta}{r^2 \nu_K} \;,
\end{equation}
where $\beta$ is the tilt angle off the equatorial plane, and $\phi_2$
is the coefficient of the quadrupole term in an expansion of the 
gravitational potential in spherical harmonics, $\phi \approx G M/r +
\phi_2 P_2(\cos\theta)$, with $P_2$ the second order Legendre 
polynomial. Outside the star, we have $\phi_2 = G A/r^3$,
where the coefficient $A$ is given, in terms of the star moment of inertia, by
$ A = (I_{xx} + I_{yy} - 2 I_{zz})/ 2$ ,
where $z$ is the star rotation axis. For rotationally flattened stars, far from
breakup, $I_{xx} = I_{yy}$.
For rotationally flattened stars, $A < 0$, and
both $I_{xx}$ and $I_{zz}$ are increased by rotation.
In the existing literature we have been able to find only the numerical 
values of $I_{zz}^{NR}$ for non-rotating neutron star models and the 
corresponding rotation--increased values, $I_{zz}^{R}$ (Friedman, Ipser and
Parker 1986, Cook, Shapiro and Teukolsky 1994). 
Using the rotating polytropic models 
calculated by James (1964), we derived the approximate relation
$I_{zz}^{NR}-I_{zz}^{R}\simeq 2(I_{xx}^{R}-I_{zz}^{R})$. Assuming that 
this holds also for neutron star models, we have 
$A \simeq (I_{zz}^{NR}-I_{zz}^{R})/2$. 
This is probably a fairly accurate approximation because the rotation 
frequencies we are interested in, $\nu_s \la 500\; Hz$, correspond to 
still slowly rotating neutron star models: 
in fact, perusal of the tabulated models in the references
given above shows that the ratio $t$ of rotational to binding energy 
for these $\nu_s$ is $\approx 0.01-0.04$, while models nearly at breakup
have $t \ga 0.11$, for all EOS. We find, defining 
$ A \equiv -\eta \nu_{s2.5}^2 I$,
where $I = I_{zz}^{NR}$, that the parameter
$\eta$ is roughly independent of spin frequency in the slow--rotation
regime of interest here, and it depends, albeit weakly, upon
model mass and EOS (range of 
$\eta \approx (0.5-3.2)\times 10^{-2}$). 
Eventually, we find
\begin{equation}
\label{nucl}
\nu_{cl} = - 4.7 I_{45} M_0^{-5/3} \left(\frac{\eta}{10^{-2}}\right) \cos\beta
\; \nu_{K3}^{7/3} \nu_{s2.5}^2 \; Hz\;.
\end{equation}
Note that the negative sign reflects the fact that classical 
precession is retrograde for $\beta < \pi/2$. 
Comparison of this with equation \ref{1} shows that, even at these small 
radii, Lense--Thirring precession is faster. The total precession 
frequency $\nu_p = \nu_{LT} + \nu_{cl}$ is thus somewhat decreased by 
oblateness--induced precession. 

Modulation of the X-ray flux at the precession frequency of the inner disk 
boundary is expected to occur through the same
geometrical effects producing the modulation at the Keplerian frequency.
In fact, the precessional motion of the blobs generating the 
signal at $\nu_K$ would give rise to a signal at the precession frequency,  
both for self-luminous blobs (the projected velocity along the line
of sight and, therefore, the Doppler boosting will be modulated at the
precession frequency) and occulting, or occulted, blobs (the duration
of the occultations wiil depend on the phase of the precessional motion).
Moreover we note that, unlike the signal at $\nu_K$, an X-ray modulation at the
precession frequency can in principle be generated, even in the absence of 
blobs, by the changing aspect of the innermost disk ring. This would modulate 
the ring emission as seen from the earth or could partially occult the 
emitting regions close to the neutron star surface.  

Lense--Thirring precession of viscous accretion disks is generally neglected
because, following the pioneering work by Bardeen and Patterson (1975),
many authors found that viscosity leads the disk to lie in the equatorial 
plane of the rotating object.  
However, this conclusion has recently been challenged by (some of) the same
authors: even if the innermost disk regions are confined to the equatorial 
plane of the neutron star, Pringle (1996) and Maloney, Begelman and Pringle 
(1996) have shown that radiation pressure can lead to 
corrugation of the disk and its lifting off the neutron star equatorial plane.
Pringle (1996) has shown that even initially planar disks are unstable
to warping, so that tilting should be generic in centrally illuminated sources.
Precessing disks have also been recently revisited by Papaloizou \etal (1997). 
It is possible that other effects, such as parametric amplification (Binney 
1978) of the
z--oscillations of diamagnetic blobs by a dipolar magnetic field tilted 
with respect to the star spin axis, may intervene (Vietri and Stella,
in preparation). 

The power spectra of 4U~1728-34 given in Fig.~1a clearly 
show a broad peak centered around a frequency of $\sim 20$, 26 and 35~Hz
when the higher frequency QPO peak is at a frequency of 
$\nu_2\sim 900$, 980 and 1100~Hz, respectively. 
The width of the $20-35$~Hz peak is of $\sim 15-20$~Hz.
The power spectrum of 4U~0614+091 (see Fig.~1b) and KS~1731-260 
(see Fig.~1c) also shows a 
similarly broad peak centered around $\sim 22$~Hz and 
$\sim 27$~Hz, when $\nu_2\sim 900$~Hz and $\nu_2\sim 1200$~Hz, respectively.
We propose that these tens of Hz QPO peaks arise from the precession
of the innermost disk regions, where the signals at the Keplerian and beat 
frequencies also originate. The thick bars in Fig.~1 represent the 
precession frequency calculated from the sum of the Lense-Thirring  
and classical frequencies (Eqs.~1 and 3) by using 
$\nu_K=\nu_2$ (see above), and $\nu_s = 363$, 323 and 262~Hz 
in the case of 4U~1728-34, 4U~0614+091 and KS~1731-260, respectively.
In all three cases, we adopted the same neutron star model with 
EOS L, $M_0=1.97$, $I_{45}/M_0 = 1.98$ and $\eta = 10^{-2}$ 
(Friedman, Ipser \& Parker 1986).  
Lense-Thirring precession dominates, with the classical term 
providing a $\sim 20$\% correction.
The fact that the (absolute) width of the tens of Hz QPO peaks
is comparable to that of the kHz QPO peaks indicates that 
lifetime broadening dominates the coherence in both cases. 

Both $\nu_K$ and $\nu_s$ are independently determined in 
these three sources. Thus the range of precession frequencies allowed 
by the model is determined {\it only} by the neutron star parameters 
$I_{45}/M_0$ and, to a lesser extent, $M_0$ and $\eta$. 
The intervals marked by the filled squares and thick bars in Fig.~1 represent 
the whole range of precession frequencies that has been obtained by using 
virtually all neutron star models given by Friedman, Ipser \& Parker (1986) 
and Cook, Shapiro \& Teukolsky (1994) 
(we excluded models with $M_0 \sim 0.5$ and ``supramassive" models).  
We regard the very good agreement with the
frequency of the corresponding power spectrum peaks 
as a strong indication in favor of our relativistic plus 
classical precession interpretation. 
We note that for all three sources the QPO peaks 
are very close to the high-frequency end of the intervals 
corresponding to the relatively stiff EOS L and $M_0\sim 2$.
As the frequency of the kHz QPOs from 4U~1728-34 varied, 
the best fit precession frequency remained in good agreement 
with the observed peak, showing that the power 
law dependence (Eq.~\ref{1}) is probably in the correct range.

The detection of kHz QPOs in the HB of GX~5-1 is
especially important for the modelling of QPOs in ``Z"-type LMXRBs. If one 
retains the standard interpretation in which the $15-50$~Hz QPOs in the 
HB arise from the MBFM, then a magnetosphere extending 
up to $> 5$ neutron star radii is required (implying a 
magnetic field of $\sim 10^{9}$~G) which argues against any interpretation 
of the kHz QPOs involving Keplerian orbits at $\sim 15-20$~km. 
If on the contrary, a Keplerian beat frequency model is adopted for the 
kHz QPOs, then the $15-50$~Hz QPOs are unlikely to arise from the MBFM. 
We regard the presence of a pair of kHz QPOs and the constancy of their 
frequency difference as strong evidence in favor of the latter
interpretation. In this context, it appears natural to interpret also 
the HB QPOs of GX~5-1 in terms of our 
precession model. For $\nu_s = \Delta\nu \sim 325$~Hz
and $\nu_K = \nu_{2} \sim 570 - 900$~Hz, the maximum precession frequency 
(again the one from EOS L and $M_0\simeq 2$) is $\sim 8-20$~Hz. 
Similarly in the case of GX~17+2, 
we obtain a maximum precession frequency of $\sim 22$~Hz, 
when the source is in the upper normal 
branch with $\nu_K = \nu_{2} \sim 988$~Hz and $\nu_s = \Delta\nu \sim 306$~Hz, 
and HB QPOs at $\sim 60$~Hz are simultaneously 
present. These precession frequencies are clearly a factor of 2--2.5 smaller 
than those of the HB QPOs. One possibility is that 
the neutron stars of ``Z"-type LMXRBs have $I_{45}/M_0\sim 4-5$, {\it i.e.}
a somewhat higher value than Friedman, Ipser \& Parker (1986) and 
Cook, Shapiro \& Teukolsky (1994) calculate. 
Alternatively the HB QPOs might arise from the 
second harmonics of the precession frequency, due to some yet unknown reason.
We note that in the case of GX~5-1 the dependence of the HB QPO frequency on 
$\nu_2$ is in the same range as the dependence of $\nu_p$ on $\nu_K$ in our 
model. 

In summary, we propose that the peaks observed in the power spectra of several
``Atoll" LMXRBs at a frequency of tens of $Hz$ (and perhaps also the 
HB QPOs of ``Z"-type LMXRBs) are due to precession of 
the innermost disk region, which is dominated by the Lense-Thirring effect.
The precession frequency depends upon the Keplerian frequency at the
innermost disk radius, and the neutron star spin frequency, both
of which are independently determined through the application of
beat frequency models to the kHz QPOs from these sources. 
The best fit is obtained 
for a stiff EOS and $M_0\simeq 2$, a value supported also from the 
interpretation of the highest observed values of $\nu_2$ 
in terms of the Keplerian frequency of the marginally stable orbit
(Kaaret, Ford \& Chen 1997; Zhang, Strohmayer \& Swank 1997).

LS acknowledges useful discussions  with D. Bini, I. Ciufolini, 
D. Psaltis. This work was partially supported through ASI grants.

\newpage

\begin{figure}
\centerline{
\psfig{figure=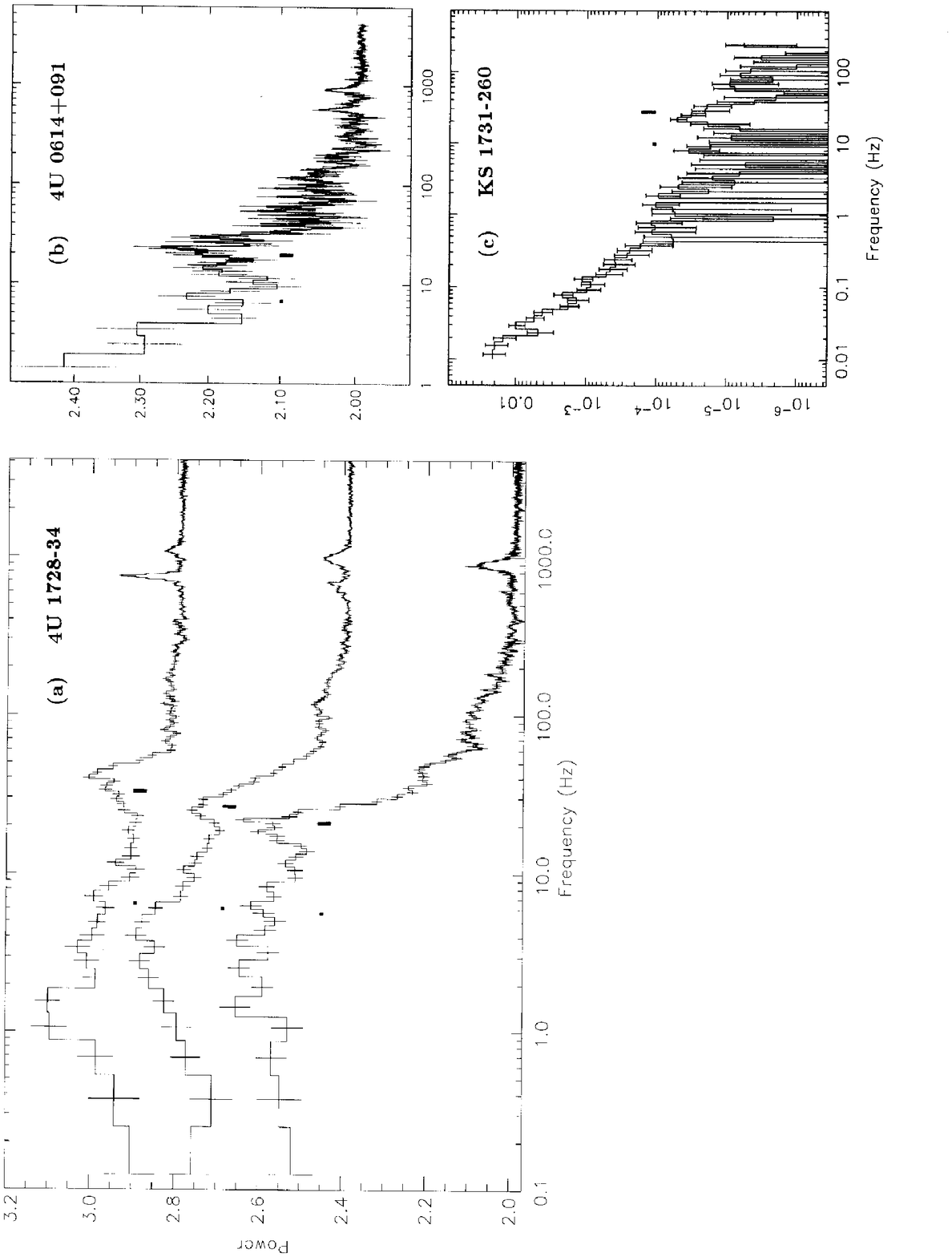,width=15cm,height=22cm}
}
\caption[]{Power spectra of the RXTE light curves of 
4U~1728-34 (panel $a$), 4U~0614+091 (panel $b$) and KS~1731-260 (panel $c$) 
(after Strohmayer \etal\/ 1996a, Ford \etal\/ 1997 and 
Wijnands \& van der Klis 1997). 
The thick bars indicate the best 
precession frequency obtained from the model discussed in the text.
The range of allowed precession frequencies for a variety of neutron 
star EOS and masses is marked by thick bars and filled squares.
\label{Figure 1}}
\end{figure}


\vskip 2truecm

\begin{thebibliography}{}

\bibitem{ } Alpar, A., \& Shaham, J. 1985, Nature 316, 239
\bibitem{ } Angelini, L., Stella, L. \& Parmar, A.N. 1989, ApJ 346, 906
\bibitem{ } Bardeen, J.M., Patterson, J.A., 1975, \apjl, 195, L65
\bibitem{ } Berger, M., \etal, 1996, ApJ, 469, L13
\bibitem{} Binney, J.J., 1978, \mnras, 183, 779
\bibitem{ } Cook, G.B., Shapiro, S.L. \& Teukolsky, S. 1994, ApJ, 424, 823
\bibitem{ } Finger, M. H., Wilson, R. B. \& Harmon, B. A. 1996, ApJ, 459, 288
\bibitem{ } Ford, E., \etal, 1997, ApJ, 475, L123
\bibitem{ } Friedman, J.L, Ipser, J.R. \& Parker, L. 1986, ApJ, 304, 115
\bibitem{ } Ghosh, P., \& Lamb, F.K. 1979, ApJ, 234, 296
\bibitem{ } Ghosh, P., \& Lamb, F.K. 1992, in {\it X-ray Binaries and 
  Recycled Pulsars}, Ed. E.P.J. van den Heuvel and S.A. Rappaport
  (Dordrecht: Kluwer), p.487
\bibitem{ } Hasinger, G., \& van der Klis, M. 1989, A\&A 225, 79
\bibitem{ } James, R. A. 1964, ApJ, 140, 552
\bibitem { } Kaaret, P., Ford, E.G. \& Chen, K. 1997, ApJL, 480, L27
\bibitem { } Kahn, F.D. \& Woltjer L. 1959, ApJ, 130, 705
\bibitem{ } Lamb, F.K., 1991, in {\it Neutron Stars: Theory and Observation}, 
  Ed. J. Ventura and D. Pines, (Dordrecht: Kluwer), p.445 
\bibitem{ } Lamb, F.K., Shibazaki, N., Alpar, A., \& Shaham, J. 1985, 
  Nature 317, 681
\bibitem{ } Lense, J., \& Thirring, H. 1918, Physik. Z. 19, 156
\bibitem{ } Maloney, P.R., Begelman, M,C., Pringle, J.E., 1996, \apj, 472, 582
\bibitem{ } Miller, M.C., Lamb, F.K., \& Psaltis, D. 1997, ApJ, submitted
\bibitem{} Papaloizou, J.C.B., Larwood, J.D., Nelson, R.P., Terquem, C., 1997,
to appear in ``Proceedings of the EARA Workshop on accretion disks",
Lecture Notes in Physics.
\bibitem{ } Pringle, J.E., 1992, \mnras, 258, 811
\bibitem{ } Pringle, J.E., 1996, \mnras, 281, 357
\bibitem{ } Smale, A.P., Zhang, W., White, N.E., 1997, ApJL, 483, L119
\bibitem{ } Smith, D.A., Morgan, E.H. \& Bradt, H. 1997, ApJ, 479, L137
\bibitem{ } Strohmayer, T., \etal, 1996a, ApJ 469, L9
\bibitem{ } Strohmayer, T., Lee, U., \& Jahoda, K. 1996b, IAU Circ. 6484
\bibitem{ } van der Klis, M. 1995, in {\it X-ray Binaries},
  Eds. W. H. G. Lewin, J. van Paradijs \& E. P. J. van den Heuvel
  (Cambridge University Press), p.~252
\bibitem{ } van der Klis, M., \etal, 1996a, IAU Circ. 6511
\bibitem{ } van der Klis, M., \etal, 1996b, ApJ, 469, L1
\bibitem{ } van der Klis, M., \etal, 1997a, IAU Circ. 6565
\bibitem{ } van der Klis, M., Wijnands, R.A.D., Horne, K. \& Chen, W. 1997b, 
   ApJL, 481, L97

\bibitem{} White, N.E. \& Zhang, W. 1997, ApJL, submitted 
\bibitem{} Wijnands, R.A.D., \etal, 1996, IAU Circ. 6447
\bibitem{ } Wijnands, R.A.D., \etal, 1997, ApJL, 479, L141 
\bibitem{ } Wijnands, R.A.D. \& van der Klis, M. 1997, ApJL, 482, L65 
\bibitem{ } Wilkins, D.C. 1972, Phys. Rev. D 5, 814
\bibitem{ } Zhang, W., Lapidus, I., White, N.E., \& Titarchuk, L. 
  1996, ApJ, 469, L17
\bibitem{ } Zhang, W., Strohmayer, T.E. \& Swank, J.H. 1997, ApJL, 482, L167
\end{thebibliography}
\end{document}